\documentclass[sigconf,nonacm]{acmart}

\usepackage{enumitem}
\usepackage{calc}
\usepackage{ragged2e}
\usepackage{multicol}
\usepackage{placeins}

\usepackage{soul}
\usepackage{microtype}
\usepackage{tabularx}
\usepackage[noend]{algpseudocode}
\usepackage{booktabs}
\usepackage{mdframed}
\usepackage{fancyvrb}
\usepackage{xcolor}
\usepackage{multirow}
\usepackage{graphicx}
\usepackage{textcomp}
\usepackage{tikz}
\usepackage{listings}
\usepackage{rotating}
\usepackage{caption}
\usepackage[linesnumbered,ruled,lined]{algorithm2e}
\usepackage{alltt}
\usepackage{framed}
\usepackage{hhline}
\usepackage{subcaption}
\usepackage[breakable]{tcolorbox}
\usepackage{float}
\usepackage{array}
\usepackage{makecell}
\usepackage{balance}
\usepackage{centernot}
\usepackage[T1]{fontenc}
\usepackage[flushleft]{threeparttable}
\usepackage{url}
\usepackage{amsthm}

\usepackage{stmaryrd}
\usepackage{lstautogobble}
\usepackage{siunitx}
\definecolor{gray}{RGB}{215,215,215}
\definecolor{light-gray}{gray}{0.8}
\definecolor{codegreen}{rgb}{0,0.6,0}
\definecolor{codegray}{rgb}{0.5,0.5,0.5}
\definecolor{mygray}{rgb}{0.7,0.7,0.7}
\definecolor{codepurple}{rgb}{0.58,0,0.82}
\definecolor{backcolour}{rgb}{0.95,0.95,0.92}
\def\BibTeX{{\rm B\kern-.05em{\sc i\kern-.025em b}\kern-.08em
    T\kern-.1667em\lower.7ex\hbox{E}\kern-.125emX}}

\newcommand{\xMapsto}[2][]{\ext@arrow 0599{\Mapstofill@}{#1}{#2}}
\def\Mapstofill@{\arrowfill@{\Mapstochar\Relbar}\Relbar\Rightarrow}

\usepackage{hyperref}
\newcommand{\issuespectertool}{\textsc{IssueSpecter}}

\newcommand{\totalissues}{10,467}

\let\othelstnumber=\thelstnumber
\def\createlinenumber#1#2{
    \edef\thelstnumber{%
        \unexpanded{%
            \ifnum#1=\value{lstnumber}\relax
              #2%
            \else}%
        \expandafter\unexpanded\expandafter{\thelstnumber\othelstnumber\fi}%
    }
    \ifx\othelstnumber=\relax\else
      \let\othelstnumber\relax
    \fi
}

\newtcbox{\mybox}[1][breakable]{on line, enlarge top by=10pt, enlarge bottom by=10pt,
     boxsep=8pt, boxrule=2pt, size=small, arc=1mm}

\definecolor{main-color}{rgb}{0.6627, 0.7176, 0.7764}
\definecolor{string-color}{rgb}{0.3333, 0.5254, 0.345}
\definecolor{key-color}{rgb}{0.8, 0.47, 0.196}
\lstdefinestyle{mystyle}
{
    language = Java,
    basicstyle = {\ttfamily \color{main-color}},
    stringstyle = {\color{string-color}},
    keywordstyle = {\color{key-color}},
    keywordstyle = [2]{\color{lime}},
    keywordstyle = [3]{\color{yellow}},
    keywordstyle = [4]{\color{teal}},
    morekeywords = [3]{<<, >>},
    morekeywords = [4]{++},
    basicstyle=\ttfamily\scriptsize,
    commentstyle=\color{blue}\ttfamily,
    morecomment=[f][\lstbg{red!20}]-,
    morecomment=[f][\lstbg{green!20}]+,
    morecomment=[f][\lstbg{yellow!20}]++,
    morecomment=[f][\lstbg{yellow!20}]--,
    morecomment=[f][\textit]{@@},
    texcl=false
}

\definecolor{grey}{rgb}{0.7,0.7,0.7}

\newcommand{\lstbg}[3][0pt]{{\fboxsep#1\colorbox{#2}{\strut #3}}}
\lstdefinelanguage{diff}{
  basicstyle=\ttfamily\scriptsize,,
  morecomment=[f][\lstbg{red!20}]-,
  morecomment=[f][\lstbg{green!20}]+,
  morecomment=[f][\lstbg{yellow!20}]++,
  morecomment=[f][\textit]{@@},
  texcl=false
}

\newcommand{\abhik}[1]{}
\newcommand{\diany}[1]{}
\newcommand{\xiang}[1]{}
\newcommand{\shinhwei}[1]{}

\newcommand{\yuanzhang}[1]{}

\newcommand{\zhuolin}[1]{}

\newcommand{\gjd}[1]{}
\newcommand{\ignore}[1]{}

\renewcommand\footnotetextcopyrightpermission[1]{}
\acmConference{}{}{}
\acmBooktitle{}
\acmDOI{}
\acmISBN{}
\setcopyright{none}

\begin{document}

\title{LLM-Guided Issue Generation from Uncovered Code Segments}

\author{Diany Pressato}
\affiliation{
  \institution{Concordia University}
  \city{Montreal}
  \country{Canada}
}
\email{diany.pressato@mail.concordia.ca}

\author{Honghao Tan}
\affiliation{
  \institution{Concordia University}
  \city{Montreal}
  \country{Canada}
}
\email{honghao.tan@mail.concordia.ca}

\author{Mariam Elmoazen}
\affiliation{
  \institution{Concordia University}
  \city{Montreal}
  \country{Canada}
}
\email{mariam.elmoazen@gmail.com}

\author{Shin Hwei Tan}
\affiliation{
  \institution{Concordia University}
  \city{Montreal}
  \country{Canada}
}
\email{shinhwei.tan@concordia.ca}

\begin{abstract}

Developers are increasingly overwhelmed by AI-generated issue reports that lack actionability and reproducibility, eroding trust in automated bug detection tools. In this paper, we present \issuespectertool{}, an automated tool that finds bugs in uncovered code segments and automatically generates prioritized, actionable issue reports. \issuespectertool{} combines coverage analysis with LLM-based defect identification, producing structured reports complete with severity ratings, reproduction steps, and suggested fixes. We evaluate  on 13 actively maintained Python projects, generating 10,467 issue reports. Manual annotation of the top-130 ranked issues by \issuespectertool{} confirms that 
84.6\% of the LLM-generated issues are valid or warrant further investigation, with 
only 15.4\% false positives. LLM-based ranking outperforms rule-based ranking by 50\% at P@3 and 41\% in MRR. The identified bugs cover a wide variety of types, from logic and boundary errors to security vulnerabilities and state consistency bugs. By ranking issues by priority, \issuespectertool{} aims to help developers focus their attention on the most impactful bugs first. Finally, we validate \issuespectertool{} through case studies reproducing real bugs surfaced from its generated issue reports, demonstrating its practical value for automatic bug discovery in open-source Python projects. Compared against CoverUp, a state-of-the-art coverage-driven test generation tool, 
\issuespectertool{} achieves a higher bug validity rate (81.0\% vs.\ 76.2\%) under 
identical evaluation conditions, using the same model and the same number of evaluated 
artifacts per project, while additionally providing structured issue reports with 
reproduction steps and candidate fixes that are immediately actionable without requiring 
developers to interpret generated test intent.


\end{abstract}



\maketitle

\pagenumbering{arabic}

\section{Introduction}

Automated Software Resolution (ASR) techniques aim to automatically fix GitHub issues, one of the most resource-intensive phases of the software development lifecycle. Recent advances in Large Language Models (LLMs) have enabled autonomous agents capable of fault localization, patch generation, and issue closure with minimal human oversight~\cite{ chen2026sweexpexperience, li2026llmbasedissue}. However, these techniques share an underlying assumption: that a human developer has already identified and submitted a bug report before any automated resolution can begin. This leaves a critical gap in the software quality assurance pipeline, as many latent defects remain undetected until they manifest as high-cost production failures.

\noindent\textbf{The Coverage-Resolution Gap.}
A fundamental gap persists between proactive testing and automated resolution. Proactive \textbf{Automated Test Generation (ATG)} tools~\cite{coverup, pynguin, schafer} increase code coverage but suffer from the \textit{oracle problem}: a generated test that reaches a buggy path often encodes the incorrect behavior into its assertions, masking the defect. Furthermore, the standard practice of filtering failing tests~\cite{pan2024swegym} systematically discards the very signals that expose validation failures. \textbf{Reactive Issue Resolution} agents~\cite{multi_agents_issue, chen2026sweexpexperience} 
possess the reasoning capacity to fix bugs but lack the \textit{activation signal} for uncovered code. Consequently, coverage-driven suites may appear comprehensive while silently masking latent vulnerabilities.

\noindent\textbf{Our Proposal: \issuespectertool{}.}
To bridge this gap, we present \issuespectertool{}, a pipeline that transforms uncovered code segments into ranked, actionable issue reports. Rather than generating tests to \textit{exercise} code, \issuespectertool{} directs LLM semantic reasoning at the \textbf{uncovered segments themselves}, where the absence of passing tests provides a precise, high-signal localization for latent defects. \issuespectertool{} operates across three stages: (i) \textit{Coverage Localization} using coverage information to identify uncovered segments, (ii) \textit{LLM-driven Defect Analysis} to generate structured reports including severity, reproduction steps, and candidate fixes; and (iii) \textit{Two-Stage Ranking} combining rule-based severity heuristics with LLM-based impact reordering to produce a prioritized triage list.

We evaluate \issuespectertool{} on 13 actively maintained open-source Python projects from the CodaMosa~\cite{codamosa} dataset. For each project, \issuespectertool{} generated and selected a 10 top-ranked issues. 
Manual annotation of these issues yielded an \textbf{84.6\% actionability rate}, with 37.7\% confirmed as genuine bugs and only 15.4\% classified as false positives. LLM-based Ranking outperforms rule-based ranking by \textbf{50\% at P@3} and \textbf{41\% in MRR}. The detected defects in these issues fall into nine of the ten categories of the adopted bug taxonomy~\cite{bugpilot} with no domain-specific tuning. Besides encountering a path traversal vulnerability (CWE-22), IssueSpecter surfaces three case studies illustrating the practical reach of these results: a memory exhaustion condition, a silent data-loss bug in a gzip decompressor, and a type constraint violation, all found in segments of fewer than 30 lines that no existing test had ever reached.

\noindent\textbf{\textit{Contributions.}} Our contributions are summarised as follows:

\begin{description}[leftmargin=*]
    \item[\textbf{Tool:}] We present \issuespectertool{}, the first automated pipeline that combines coverage-guided segment identification with LLM-based defect detection to generate prioritized, structured issue reports from uncovered code segments in Python projects.
    \item[\textbf{Study:}] We conduct an evaluation on 13 open-source Python projects, generating 
10,467 issue reports and manually annotating 130 top-ranked issues, showing that 84.6\% 
are valid or warrant further investigation. We further compare \issuespectertool{} 
against CoverUp over 168 matched artifacts per tool, with \issuespectertool{} achieving 
a higher bug validity rate (81.0\% vs.\ 76.2\%), with LLM-based Ranking consistently 
outperforming rule-based ranking across all evaluated metrics.
    \item[\textbf{Case Studies:}] We validate \issuespectertool{} through case studies reproducing real bugs across diverse bug types, demonstrating its practical value for surfacing bugs in open-source Python projects.
\end{description}
\section{Motivating Example}
\label{section:motivating_example}

To explain the design of various components of \issuespectertool{}, consider a concrete scenario with HTTPie\footnote{https://github.com/httpie/cli}, a widely used HTTP client library, having 2,123 source files and a test coverage of 32.2\%, with 206 uncovered code segments present in the project. These uncovered segments represent self contained blocks of untested logic which may contains bugs that are overlooked during software development. 
After identifying these uncovered segments,  our Issue Generation phase, \issuespectertool{} produces 618 issue reports in total, of which 592 identify potential bugs either in the code or in its documentation. 63 of these generated issues are flagged as having high severity, a number that could be time-consuming for a human developer to manually inspect and verify their correctness.

Without an automated selection step, a developer might overlook 
critical issues such as a path traversal vulnerability (CWE-22), 
confirmed through manual reproduction, which remains buried among hundreds 
of candidates. As this vulnerability was found in a short uncovered segments, this shows that even short untested parts of the code may contain serious bugs.

\issuespectertool{} addresses this through its two-stage reduction pipeline: a rule-based Repository-level Issue Selection phase that narrows the 618 reports down to the top 10 most important issues per project, followed by an LLM-based Ranking phase that reorders those 10 by impact, scope, and urgency. For the HTTPie project, the LLM-based Ranking achieves a perfect MRR score of 1.00, compared to 0.14 under rule-based ranking alone, placing the highest-priority issue (represented by a path traversal vulnerability), at the top of the ranked list. This example illustrates why ranking is necessary, since  without it, the most critical bugs risk are probably being overlooked, even in a project with a substantial test suite of 1,028 unit tests.

Table~\ref{tab:httpie-ranking} shows the top 10 issues ranked by the rule-based sorting criteria for the HTTPie project, and the rankings produced by the LLM and the human annotators. The rule-based approach fails to rank because all 10 top-selected HTTPie issues share the same severity level and OS score, as shown in Table~\ref{tab:httpie-ranking}.  
As all candidates share the same ranking based on the severity and OS score, the rule-based ranking degenerates into a simple word count ordering that is agnostic to the actual impact of each defect. As a result, the path traversal vulnerability in function \texttt{session\_hostname\_to\_dirname} (CWE-22) is ranked $7^{th}$, because its description is shorter than other issues. 
Meanwhile, the LLM-based Ranking correctly elevates this security vulnerability to position 1, in agreement with the golden ranking produced by human annotators, while also correctly attributes low priority to issues such as the \texttt{output\_file\_specified}, demonstrating that LLM-based Ranking can help further improve the ranking given by the rule-based ranking phase. 

\begin{table}[h]
\centering
\footnotesize
\setlength{\tabcolsep}{2pt}
\caption{Top-4 ranking comparison for the HTTPie project. All issues share a Critical Priority Score and affect all Operating Systems; the Rule-based Ranking and the LLM-based Ranking are compared against the Golden Ranking produced by human annotators.}
\label{tab:httpie-ranking}
\begin{tabular}{p{3.5cm}ccc c}
\toprule
\textbf{Issue Title} & \textbf{Word} & \textbf{Golden} & \textbf{Rule-based} & \textbf{LLM-based} \\
 & \textbf{Count} & \textbf{Ranking} & \textbf{Ranking} & \textbf{Ranking} \\
\midrule
session\_hostname\_to\_dirname: unsanitized inputs leading to path traversal & 444 & 1 & 7 & 1 \\
\addlinespace
parse\_content\_type\_header wrongly splits on semicolons in quoted values & 531 & 2 & 1 & 4 \\
\addlinespace
Unbounded buffering in BufferedPrettyStream.iter\_body can cause OOM & 447 & 3 & 6 & 3 \\
\addlinespace
Fallback to localhost for hostless URLs causes session collisions & 459 & 4 & 4 & 2 \\
\bottomrule
\end{tabular}
\end{table}
\section{Methodology}

\begin{figure}[ht]
    \centering
    \includegraphics[width=0.7\linewidth]{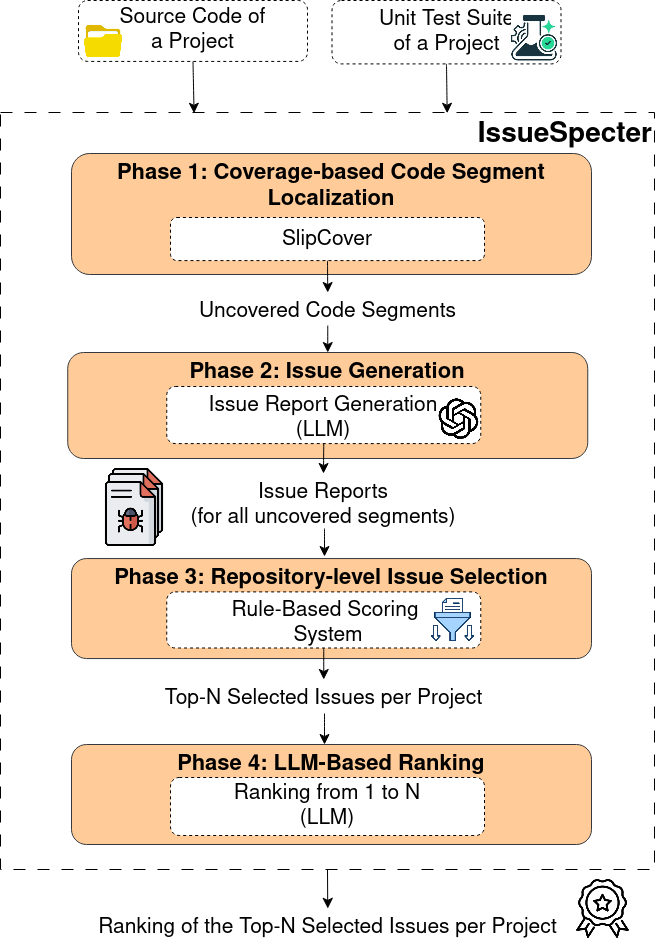}
    \caption{\issuespectertool{}'s overall workflow.}
    \label{issuespecter_pipeline}
\end{figure}



%

Figure~\ref{issuespecter_pipeline} shows \issuespectertool{}'s overall workflow. It automatically generates and prioritizes issue reports via a four-phase approach: (1) Coverage-based Code Segment Localization, (2) Issue Generation, (3) Repository-level Issue Selection, (4) LLM-based Ranking. 

\noindent\textbf{Coverage-based Code Segment Localization:} Instead of generating issues for all the source code within a given project, our key insight in issue generation is that  \emph{uncovered code segments correspond to untested segments that are more likely to contain bugs}. Given the source code and the corresponding unit test suite of a project, \issuespectertool{} automatically identifies code segments that are uncovered by existing unit test suite.  Specifically,
we use SlipCover~\cite{slipcover}, a code coverage tool for Python that can compute a set of uncovered code with minimal overhead. 

\noindent\textbf{Issue Generation:} 
For each uncovered segment identified in previous phase, \issuespectertool{} uses GPT-5-mini to identify potential defects by prompting the model to discover up to three distinct bugs per each uncovered segment and generate corresponding issue reports. Figure~\ref{fig:bug_prompt_template} shows our prompt template used for automated bug identification. It specifies that every generated issue should contain: a severity category (Critical, High, Medium, Low, or Very Low) assigned by the LLM, the affected operating systems, an in-depth description with step-by-step reproduction instructions, and suggested code fixes for the identified defect.

\noindent\begin{minipage}{\linewidth}
\begin{tcolorbox}[
    width=0.95\linewidth,
    colback=gray!5,
    coltext=black,
    colframe=black!60,
    boxrule=0.6pt,
    arc=2pt,
    left=6pt,
    right=6pt,
    top=6pt,
    bottom=6pt
]
\footnotesize

\textbf{Role:} You are an expert Python developer and test engineer. 
You will be provided with a Python code snippet extracted from \texttt{{PROJECT\_NAME}}.

\textbf{Task:}
Analyze the code, identify up to three distinct defects, and generate corresponding 
bug reports. In the absence of bugs, include a boolean indicator explicitly stating 
that no bug was found.

For each bug you identify, generate an issue report and provide an associated 
pull request suggestion. After completing the issue report, organize all found 
issues according to the precise JSON schema output. For every proposed fix, ensure 
the code maintains original functionality and style.

For each bug, assess its severity (very low, low, medium, high, or critical) and 
specify affected operating system(s). Indicate whether documentation inconsistencies 
are involved and propose a Python code fix when appropriate, preserving original 
functionality and coding style.

\textbf{Input:}
Python code snippet to be analyzed: \texttt{{CODE\_SNIPPET}}

\textbf{JSON Output:}
A JSON array of three objects, each with a boolean indicating whether a bug is 
present. If a bug exists, the object includes title, summary, bug severity, OS, 
generated issue, a boolean for inconsistent documentation, and fixed code, ensuring 
consistent, parseable output.

\end{tcolorbox}
\captionof{figure}{Prompt template used for automated bug identification and reporting during our Issue Generation phase.}
\label{fig:bug_prompt_template}
\end{minipage}

\vspace{0.2cm}
\noindent\textbf{Repository-level Issue Selection:} 
The Issue Generation phase may produce a hundreds of issue reports across a project's uncovered code segments. Hence, the goal of this phase is to select a list of important issues considering all uncovered code segments within a project. Specifically, we design a rule-based ranking approach based on prior research on bug reports prioritization ~\cite{preranking_criteria}. 

Our selection strategy employs a three-steps hierarchical ranking mechanism, as summarized in Table~\ref{tab:scoring_criteria}. Issues are ranked according to the following priority order:

\noindent\textbf{(1) Bug Severity Priority:} Issues are first grouped into five severity levels, with higher severity groups taking absolute precedence. 

\noindent\textbf{(2) Operating System Impact:} Within each severity group, issues are then sorted by the number of affected operating systems. Issues affecting all platforms receive the highest OS score of 100, while others receive a score equal to the number of distinct OS labels (e.g., 2 for an issue affecting Windows and macOS). Issues with higher OS scores are ranked first. 

\noindent\textbf{(3) Description Completeness:} Finally, among issues with identical severity and OS scores, we rank based on the word count, using description length as a proxy for completeness under the assumption that longer issues contain more detailed reproduction steps and context.

This hierarchical ranking ensures that a critical bug affecting one platform is prioritized over a low-severity bug affecting all platforms, and that cross-platform issues within the same severity level are ranked above single-platform issues. 
For each project, this heuristic ranking selects top 10 issues, representing a list of high-severity issues with detailed descriptions.


%


\FloatBarrier
\begin{table}[htbp]
\centering
\caption{Three-steps hierarchical ranking criteria for Repository-level Issue Selection.}
\label{tab:scoring_criteria}
\small
\begin{tabular}{@{}clp{4.5cm}@{}}
\toprule
\textbf{Order} & \textbf{Criterion} & \textbf{Description} \\
\midrule
1 & Priority   & \raggedright Severity grouping: Critical, High, Medium, Low and Very Low.\arraybackslash \\
2 & OS Score   & \raggedright 100 for ``all''; else distinct OS count (higher = more platforms).\arraybackslash \\
3 & Word Count & \raggedright Description length (higher = more complete).\arraybackslash \\
\bottomrule
\end{tabular}
\end{table}


\noindent\textbf{LLM-based Ranking:} The fourth phase employs an LLM-based Ranking, in which GPT-5-mini evaluates each project's top 10 issues selected by the rule-based selection, and then assigns a priority ranking from 1 (highest priority) to 10 (lowest priority) for each issue according to \textit{impact}, \textit{scope} and \textit{urgency}, as illustrated in the prompt template in Figure ~\ref{fig:issue_ranking_prompt}. \issuespectertool{} produces two outputs: 
(1) the rule-based ranking derived from Phase 3, and (2) the LLM ranking for the same set of top-selected issues per project, where the ranking process is performed independently for each project.

\noindent\begin{minipage}{\linewidth}
\begin{tcolorbox}[
    width=0.95\linewidth,
    colback=gray!5,
    coltext=black,
    colframe=black!60,
    boxrule=0.6pt,
    arc=2pt,
    left=6pt,
    right=6pt,
    top=6pt,
    bottom=6pt
]
\footnotesize

\textbf{Role:}
You are a software engineering expert tasked with ranking up to 10 issue reports 
from a specified project according to urgency, scope and bug impact.

\textbf{Task:}
Each issue includes associated attributes: \textit{bug severity}, \textit{affected os}, 
\textit{number of unit tests failing} from the project test suite when the proposed 
Python fix is applied, and \textit{word count}.

Evaluate each issue based on logical validity, alignment with best practices and 
documentation, and support from external technical resources, when available, and 
assess the issue's fit with the project's architecture and design patterns. 
Additionally, evaluate the impact of failed tests, noting if a proposed fix passes 
all tests. Rank issues in descending order of impact and urgency, with ``1'' as the 
highest. If required data is missing or unverifiable, clearly indicate this in the 
output.

\textbf{JSON Output:}
For each ranked issue, generate a structured Markdown report under the 
\textit{validity report} field. The final output should be a valid JSON object 
with the ranking of each issue, with each object containing \textit{issue id}, 
\textit{reasoning}, \textit{validity classification} (a boolean to indicate if 
the issue is valid or not), \textit{confidence rating}, and the 
\textit{validity report}.

\textbf{Input JSON:} The JSON object input to process:
\texttt{{ISSUES\_ATTRIBUTES}}

\end{tcolorbox}
\captionof{figure}{Prompt template used for LLM-based Ranking.}
\label{fig:issue_ranking_prompt}
\end{minipage}

\noindent\textbf{Patch correctness.} We use existing test suite to further rank the issues based on patch correctness. For each original program $P$ and the fixed program $P'$ (included in the issue), we execute all tests within the given test suite to identify regression-introduced patches by checking whether there are new failing tests (i.e., tests that passes in $P$ but fails in $P'$). Issues with regression-introduced patches are ranked lower, with the ranking decreasing further as the number of failing tests increases for a given patch.

\FloatBarrier
\section{Evaluation}
\label{sec:evaluation}


Our evaluation focuses on three research questions:

\noindent\textbf{RQ1:} How effective is \issuespectertool{} in detecting bugs in uncovered code segments?

\noindent\textbf{RQ2:} How do the individual components of \issuespectertool{} contribute to its overall effectiveness?

\noindent\textbf{RQ3:} How does \issuespectertool{} compare with unit test generation approaches in surfacing bugs?



\FloatBarrier
\subsection{Dataset}

Our evaluation consists of a total of 13 open-source projects, as shown in Table~\ref{tab:project_stats2}.
Specifically, from the 27 projects in the CodaMosa dataset~\cite{codamosa}, we selected project that are actively maintained (e.g, with commits after 2023), excluding 11 inactive projects. Active maintenance is a critical selection criterion for our study, as we generate issues for the latest version of the project, validate generated bug reports by submitting them as issues and pull requests to the respective repositories, enabling us to obtain future community feedback. In addition, three projects with recent activity were excluded due to technical incompatibilities with SlipCover during coverage analysis (i.e., dependency conflicts that prevented the extraction of uncovered code segments). 
Overall, the 13 selected projects covers diverse domains that encompasses developer tools, web frameworks, data utilities, automation systems, and multimedia applications, ensuring \issuespectertool{} captures bugs across different programming paradigms. 
The selected projects are diverse in terms of complexity and test coverage, with project sizes varying considerably, ranging from compact codebases like TQDM with 67 files to bigger ones, such as iSort with 5,049 files. 
We focus on widely used Python projects that have been used for prior evaluation of unit test generation techniques~\cite{codamosa}, \cite{coverup}, \cite{pynguin}, \cite{mutap}, \cite{testgenllm}, \cite{schafer}, \cite{unit_test_cov}, \cite{copilot_unit_test_py}, \cite{pynguin_type_inf}, \cite{gen_ai_python}, \cite{gpt_pynguin_units_py}, \cite{pynguin_extension}, \cite{chatgpt_units}.

\begin{table}[t]
\caption{Characteristics of the 13 selected Python projects from CodaMosa dataset. Metrics include number of files, mean and maximum lines of code per file, test coverage percentage, commit hash at analysis time, and number of existing tests.}
\label{tab:project_stats2}
\small
\setlength{\tabcolsep}{3pt}
\renewcommand{\arraystretch}{0.9}
\centering
\begin{tabular}{lrrrrrr}
\toprule
Project & \#Files & Mean & Max & Cov. & Hash & \#Tests \\
\midrule
Ansible~\cite{ansible} & 1,573 & 94.5 & 2,235 & 46.2\% & 22721 & 2,062 \\
Cookiecutter~\cite{cookiecutter} & 89 & 70.5 & 590 & 88.6\% & af1d7 & 382 \\
Dataclasses-JSON~\cite{dataclasses_json} & 39 & 94.9 & 358 & 94.8\% & dc639 & 324 \\
DocstringParser~\cite{docstring_parser} & 1,001 & 277.5 & 8,510 & 94.7\% & 43407 & 230 \\
HTTPie~\cite{httpie} & 2,123 & 211.8 & 8,664 & 32.2\% & 5b604 & 1,028 \\
isort~\cite{isort} & 5,049 & 215.1 & 20,748 & 16.0\% & 77bc6 & 536 \\
Mimesis~\cite{mimesis} & 90 & 577.0 & 13,068 & 99.1\% & 70b11 & 8,268 \\
PySnooper~\cite{pysnooper} & 1,013 & 147.7 & 5,158 & 40.1\% & 0561a & 78 \\
Sanic~\cite{sanic} & 2,860 & 214.4 & 8,664 & 96.2\% & a64d7 & 1,693 \\
Thonny~\cite{thonny} & 328 & 184.4 & 3,432 & 37.3\% & 55ac2 & 3 \\
Tornado~\cite{tornado} & 111 & 258.2 & 2,560 & 30.7\% & d30ef & 1,317 \\
tqdm~\cite{tqdm} & 67 & 85.5 & 1,427 & 71.0\% & 0ed5d & 150 \\
Youtube-DL~\cite{youtube_dl} & 902 & 152.3 & 5,158 & 62.5\% & 1e109 & 2,767 \\
\midrule
Total & 15,245 & --- & --- & --- & --- & 18,838 \\
Mean & 1,173 & 198 & 6,120 & 62.3\% & --- & 1,449 \\
\bottomrule
\end{tabular}
\end{table}

\subsection{Segments}

Table~\ref{tab:uncovered_segments} shows the uncovered segments that \issuespectertool{} selected as targets for issue generation. The evaluated projects have an average of 15 to 25 lines per segment, suggesting that the \emph{uncovered regions tend to be compact}, representing self-contained blocks, rather than large untested modules.


\begin{table}[htpb]
  \centering
  \small
  \caption{For each project, we report the total number of uncovered segments identified by SlipCover, along with the maximum, minimum, and average number of source lines spanned by those segments.}
  \label{tab:uncovered_segments}
  \begin{tabular}{lrrrr}
    \toprule
    \textbf{Project}
      & \thead{\textbf{\# Uncovered} \\ \textbf{Segments}}
      & \thead{\textbf{Lines} \\ \textbf{Maximum}}
      & \thead{\textbf{Lines} \\ \textbf{Minimum}}
      & \thead{\textbf{Lines} \\ \textbf{Average}} \\
    \midrule
    Ansible           & 1832 & 364 &  1 & 20.77 \\
    Cookiecutter      &   25 & 101 &  2 & 24.32 \\
    Dataclasses-JSON  &   55 &  84 &  1 & 17.40 \\
    Docstring Parser  &   51 & 151 &  1 & 23.78 \\
    HTTPie            &  206 & 111 &  1 & 15.09 \\
    iSort             &   24 &  30 &  4 & 10.83 \\
    Mimesis           &  300 &  63 &  1 & 11.27 \\
    PySnooper         &   35 & 176 &  1 & 18.14 \\
    Sanic             &  322 & 188 &  1 & 21.98 \\
    Thonny            &   28 & 135 &  2 & 25.46 \\
    Tornado           &  284 & 193 &  1 & 18.35 \\
    tqdm              &   44 &  59 &  1 & 16.09 \\
    Youtube-DL        &  283 & 901 &  1 & 26.00 \\
    \bottomrule
  \end{tabular}
\end{table}

\section{RQ1: Effectiveness of \issuespectertool{}}
\label{sec:rq1}

\paragraph{Metrics.}
We evaluate the effectiveness of \issuespectertool{} by manually annotating the top-ranked issues and measuring the \emph{error detection rate}: the proportion of generated issues that are either confirmed bugs or plausible defects warranting further investigation. We also characterize the diversity of surfaced defect types using the ten-category bug taxonomy from~\cite{bugpilot}.

\paragraph{Annotation Procedure.}
Two authors independently read each of the top-10 ranked issues per project (130 issues in total across 13 projects) and classified each as one of: \emph{Valid Bug} (high confidence the defect exists), \emph{Requires Further Investigation} (plausible but not fully reproducible, e.g., edge cases), or \emph{Invalid} (hallucination or misunderstanding of intended usage). For each issue, annotators also assigned a single bug category from the ten-category taxonomy of~\cite{bugpilot}, with the annotated bug distribution shown in Table~\ref{tab:bug_distribution}. Note that this annotation covers the 130 issues used to evaluate \issuespectertool{}'s 
effectiveness (RQ1). The comparison against CoverUp in Section~\ref{sec:rq3} is 
conducted over a separate set of 168 matched artifacts per tool, to ensure a fair comparison and following the same 
classification scheme.

\paragraph{Issue Generation Volume.}

The \totalissues{} issues generated by \issuespectertool{} across the 13 selected projects, covers five severity levels (Critical, High, Medium, Low and Very Low). The Issue Generation phase achieved a high error detection rate: 9,929 out of \totalissues{} issues, corresponding to 94.9\% of the issues, are flagged with at least one actual defect in the analyzed uncovered segment. This rate should be interpreted with caution rather than taken as a direct measure of bug detection accuracy, since LLMs tend to exhibit confirmation bias when prompted to find bugs (i.e., a model may report a defect even in correct code).

\paragraph{Annotation Results.}
Table~\ref{tab:annotation_results} shows the results of the annotation, with an agreement of 80.3\%. 84.6\% of top-ranked issues are either confirmed valid bugs (37.7\%) or require further investigation (46.9\%), with only 15.4\% classified as invalid (e.g., hallucinations or misunderstandings of the project's intended usage patterns). Although a relatively high percentage of the top-10 issues selected by our rule-based ranking approach are valid bugs, we believe that exposing developers to a non-trivial rate of invalid reports risks eroding trust in LLM-assisted tools over time, reinforcing the importance of the selection and ranking phases of \issuespectertool{} as a necessary layer between raw LLM output and developer attention.

\begin{table}[H]
  \centering
  \caption{Manual assessment of 130 top-ranked generated issues by two 
  human annotators. Valid bugs exhibit high confidence of existence; 
  issues requiring further investigation appear plausible but require 
  execution for confirmation or represent edge cases; invalid issues are 
  hallucinations or misunderstandings of project usage.}
  \label{tab:annotation_results}
  \begin{tabular}{lrr}
    \toprule
    \textbf{Bug Assessment} & \textbf{Count} & \textbf{Percentage} \\
    \midrule
    Valid Bug                       & 49  & 37.7\% \\
    Requires Further Investigation  & 61  & 46.9\% \\
    Invalid                         & 20  & 15.4\% \\
    \midrule
    \textbf{Total}                  & \textbf{130} & \textbf{100\%} \\
    \bottomrule
  \end{tabular}
\end{table}

\paragraph{Bug Type Diversity.}
Table~\ref{tab:bug_distribution} shows that the validated issues skew heavily toward Logic and Conditional Bugs (43 cases, 33.1\%) and Input Validation, Boundary, and Sentinel Handling Errors (39 cases, 30.0\%).
Together, these categories make up over 63\% of the total annotated issues, suggesting that LLMs are particularly sensitive to control-flow and boundary-handling problems residing in uncovered code, precisely the category of defects that test suites most commonly fail to exercise.
Security-related defects (Security and Sensitive Data Leakage Due to Logic Oversight) were less frequent but nonetheless present, with four identified cases.

\begin{table}[ht]
\centering
\caption{Distribution of bug categories across validated issues generated by GPT-5-mini, manually annotated according to the taxonomy from \cite{bugpilot}. Logic errors and input validation issues dominate, while I/O handling and mutability bugs are less frequently detected for our 130 validated issues.}
\label{tab:bug_distribution}
\small
\begin{tabular}{@{}p{7.0cm}r@{}}
\toprule
\textbf{Bug Category} & \textbf{Count} \\
\midrule
Logic and Conditional Bug & 43 \\
Input Validation, Boundary, and Sentinel Handling Error & 39 \\
State Consistency, Bookkeeping, and Caching Bug & 14 \\
Protocol and Specification Conformance Bug & 12 \\
API, Signature Mismatch, or Backward Compatibility Break & 10 \\
Incorrect Argument Forwarding, Constructor, or Inheritance Contract Break & 6 \\
Security and Sensitive Data Leakage Due to Logic Oversight & 4 \\
IO, Filesystem, or Resource Handling Bug & 1 \\
Copy Semantics, Mutability Aliasing, or In Place Mutation of Inputs & 1 \\
Missing import / symbol / attribute error & 0 \\
\midrule
\textbf{Total} & \textbf{130} \\
\bottomrule
\end{tabular}
\end{table}

The distribution of surfaced defects spans nine out of ten categories of the adopted taxonomy, with no single category accounting for more than one third of all validated issues.
This diversity reflects the breadth of Python defect patterns that LLM-based analysis of uncovered segments is capable of exposing, spanning from low-level type mismatches and boundary errors to higher-level security vulnerabilities and protocol conformance violations.

Domain-agnostic patterns hold consistently across web frameworks (Sanic, Tornado), developer tools (PySnooper, Thonny), data utilities (Mimesis, Dataclasses-JSON), and automation systems (Ansible), suggesting that logic and boundary errors represent a fundamental class of defects that systematically escape test coverage in Python projects.
Domain-specific observations emerge at finer granularity: security-related defects were exclusively found in projects that handle external input or perform filesystem and network operations; state consistency and bookkeeping bugs were disproportionately present in projects with asynchronous or streaming architectures; and API and signature mismatch bugs were most prevalent in large, multi-component projects such as Ansible.

\begin{tcolorbox}[left=0pt,right=0pt,top=0pt,bottom=0pt]
\textbf{Finding 1:} \issuespectertool{} generates top-ranked issues 
of which 84.6\% are either confirmed valid bugs (37.7\%) or require 
further investigation (46.9\%), with only 15.4\% classified as 
outright false positives. Our defect distribution suggests that 
LLM-based analysis of uncovered code segments can be effective at 
identifying control-flow and boundary-handling defects.

\textbf{Implication 1:} \issuespectertool{} issue generation from uncovered code segments could be a practical approach to complement existing test suites by systematically ranking potential bugs, spanning a set of diverse bug types, that may warrant further investigation.
\end{tcolorbox}

\section{RQ2: Ablation Study}
\label{sec:rq2}

\paragraph{Annotation Setup.} To evaluate the effectiveness of both rule-based and LLM-based Ranking, two annotators independently reviewed the top 10 issues selected per project, totaling 130 issues across 13 projects, with an annotation agreement of 82\%. Each issue was classified according to the bug taxonomy from BugPilot~\citep{bugpilot} and assessed for bug validity as shown in Table~\ref{tab:bug_distribution}.

Annotators also produced a golden ranking of each project's 10 issues from 1 (highest priority) to 10 (lowest priority), using the same three criteria employed by the LLM-based ranker: \emph{impact} (severity of effect on functionality or users), \emph{scope} (breadth of affected users, features, or components), and \emph{urgency} (prioritizing security vulnerabilities and regressions).
These annotated rankings serve as ground truth for evaluating both ranking strategies.

\paragraph{Metrics.} We evaluate both ranking strategies using the following metrics:

\noindent\emph{Precision at $k$ (P@$k$).} Measures the fraction of relevant 
issues among the top-$k$ ranked results, rewarding strategies that surface true positives early in the ranking.

\noindent\emph{Normalized Discounted Cumulative Gain (NDCG@$k$).} Measures 
ranking quality by assigning higher weights to relevant results appearing at 
higher ranks, normalized against an ideal ranking:
\begin{equation}
    \text{NDCG@}k = \frac{\text{DCG@}k}{\text{IDCG@}k}, \quad 
    \text{where DCG@}k = \sum_{i=1}^{k} \frac{\text{rel}_i}{\log_2(i+1)}
\end{equation}

\noindent\emph{Mean Reciprocal Rank (MRR).} Measures the average reciprocal 
rank of the first relevant result across all queries:
\begin{equation}
    \text{MRR} = \frac{1}{|Q|} \sum_{q=1}^{|Q|} \frac{1}{\text{rank}_q}
\end{equation}

\noindent\emph{Expected Reciprocal Rank (ERR).} Extends MRR by modeling user 
satisfaction as a cascade, where the probability of examining a result depends 
on the relevance of all preceding results.

\noindent\emph{Mean Average Precision (MAP).} Computes the mean of Average 
Precision scores across all queries, where Average Precision rewards rankings 
that place all relevant results as high as possible:
\begin{equation}
    \text{MAP} = \frac{1}{|Q|} \sum_{q=1}^{|Q|} \text{AP}(q), \quad
    \text{where AP}(q) = \frac{1}{R}\sum_{k=1}^{n} P@k \cdot \mathbb{1}[d_k \in \mathcal{R}]
\end{equation}

\paragraph{Precision Results.}
Tables~\ref{tab:precision} and~\ref{tab:ndcg} present the comparison of rule-based versus LLM-based Ranking.
LLM-based Ranking achieves higher precision at all cutoff points.
At P@1, LLM-based Ranking correctly places the highest-priority issue in 4 projects, compared to only 2 for rule-based ranking, showing a 100\% improvement at the position most relevant for developers who can inspect only one issue.
This advantage increases at P@3, where LLM-based Ranking achieves a cumulative score of 8.00, while rule-based ranking scores 5.33, representing a 50\% improvement.
At P@5, LLM-based Ranking maintains a consistent 19\% improvement (8.60 vs.\ 7.20).
Notable individual improvements include HTTPie, which improves P@1 from 0 to a perfect score under LLM-based Ranking and achieves perfect precision at P@5, and Youtube-DL, which reaches a perfect P@1 under LLM-based Ranking compared to 0 in rule-based ranking.

\paragraph{\textbf{NDCG Results.}} LLM-based Ranking shows higher cumulative NDCG at every cutoff point (Table~\ref{tab:ndcg}). NDCG rewards relevance while increasingly penalizing relevant items placed lower in the ranking.
At NDCG@1, LLM-based Ranking scores 9.90, compared to 9.40 for rule-based ranking.
This gap widens progressively, reaching 11.98 for LLM-based versus 11.61 for rule-based at NDCG@10.
The consistent improvements at NDCG@10 across most projects suggest that LLM-based Ranking performs better across the entire top-10 list, not just at the top positions.

\paragraph{MRR, ERR, and MAP Results.}
As shown in Table~\ref{tab:other_metrics}, LLM-based Ranking achieves a cumulative MRR of 7.21, compared to 5.11 for rule-based ranking, showing a 41\% improvement.
HTTPie improves from MRR\,=\,0.14 to 1.00, and Youtube-DL from 0.25 to 1.00.
For ERR, LLM-based Ranking scores 8.83 vs.\ 7.22 for rule-based (22\% improvement), with significant gains in HTTPie, Youtube-DL, and PySnooper.
MAP improves from 8.05 to 9.33 (16\% improvement). 

Despite these consistent improvements, accurately retrieving the single most relevant issue remains challenging for both approaches.
However, LLM-based Ranking successfully reaches 66\% of projects at P@5, offering meaningful practical value over heuristics alone.

\begin{table}[H]
\centering
\small
\caption{Precision at k comparison between the rule based approach and LLM-based Ranking across 13 projects. Values represent the proportion of valid or potential bugs among the top-k ranked issues. Bold indicates superior performance.}
\label{tab:precision}
\begin{tabular}{l|ccc|ccc}
\hline
\multirow{2}{*}{\textbf{Project}} & \multicolumn{3}{c|}{\textbf{Rule-Based Ranking}} & \multicolumn{3}{c}{\textbf{LLM-based Ranking}} \\
& \textbf{P@1} & \textbf{P@3} & \textbf{P@5} & \textbf{P@1} & \textbf{P@3} & \textbf{P@5} \\
\hline
ansible & 1.00 & 0.67 & 0.60 & 1.00 & \textbf{1.00} & \textbf{0.80} \\
cookiecutter & 0.00 & 0.33 & 0.60 & 0.00 & \textbf{0.67} & \textbf{0.80} \\
dataclasses\_json & 1.00 & 0.33 & 0.60 & 1.00 & \textbf{1.00} & \textbf{0.80} \\
docstring\_parser & 0.00 & \textbf{0.67} & 0.60 & 0.00 & 0.33 & 0.60 \\
httpie & 0.00 & 0.33 & 0.60 & \textbf{1.00} & \textbf{0.67} & \textbf{1.00} \\
isort & 0.00 & 0.00 & 0.60 & 0.00 & \textbf{0.33} & 0.60 \\
mimesis & 0.00 & 0.33 & 0.40 & 0.00 & 0.33 & 0.40 \\
pysnooper & 0.00 & 0.67 & 0.40 & 0.00 & 0.67 & \textbf{0.60} \\
sanic & 0.00 & 0.67 & 0.80 & 0.00 & \textbf{1.00} & 0.80 \\
thonny & 0.00 & 0.33 & 0.40 & 0.00 & 0.33 & 0.40 \\
tqdm & 0.00 & 0.33 & 0.40 & 0.00 & 0.33 & 0.40 \\
tornado & 0.00 & 0.33 & 0.60 & 0.00 & \textbf{0.67} & \textbf{0.80} \\
youtube\_dl & 0.00 & 0.33 & 0.60 & \textbf{1.00} & \textbf{0.67} & 0.60 \\
\hline
\textbf{Total} & 2.00 & 5.32 & 7.20 & \textbf{4.00} & \textbf{8.00} & \textbf{8.60} \\
\hline
\end{tabular}
\end{table}

\begin{table}[h]
\centering
\small
\setlength{\tabcolsep}{4pt}
\caption{NDCG metrics comparison between Rule Based and LLM ranking at cutoffs @1, @3, and @5.}
\begin{tabular}{l|ccc|ccc}
\hline
\multirow{2}{*}{Project} & \multicolumn{3}{c|}{Rule Based} & \multicolumn{3}{c}{LLM} \\
& @1 & @3 & @5 & @1 & @3 & @5 \\
\hline
ansible         & 1.00 & 0.94 & 0.85 & 1.00 & \textbf{1.00} & \textbf{0.97} \\
cookiecutter    & 0.10 & 0.53 & 0.65 & \textbf{0.60} & \textbf{0.85} & \textbf{0.87} \\
dataclasses\_json & 1.00 & 0.72 & 0.73 & 1.00 & \textbf{1.00} & \textbf{0.98} \\
docstring\_parser & 0.90 & \textbf{0.82} & 0.81 & 0.90 & 0.80 & 0.81 \\
httpie          & 0.40 & 0.65 & 0.73 & \textbf{1.00} & \textbf{0.94} & \textbf{0.98} \\
isort           & 0.60 & 0.51 & 0.66 & \textbf{0.80} & 0.51 & \textbf{0.67} \\
mimesis         & 0.70 & 0.77 & 0.72 & 0.70 & 0.77 & 0.72 \\
pysnooper       & 0.90 & 0.72 & 0.63 & 0.90 & \textbf{0.83} & \textbf{0.85} \\
sanic           & \textbf{0.50} & \textbf{0.80} & \textbf{0.88} & 0.20 & 0.65 & 0.76 \\
thonny          & \textbf{0.70} & \textbf{0.77} & \textbf{0.69} & 0.10 & 0.50 & 0.57 \\
tqdm            & 0.90 & 0.81 & 0.74 & 0.90 & 0.81 & 0.74 \\
tornado         & 0.70 & 0.71 & 0.67 & \textbf{0.80} & \textbf{0.89} & \textbf{0.85} \\
youtube\_dl     & 0.70 & 0.78 & 0.79 & \textbf{1.00} & \textbf{0.94} & \textbf{0.89} \\
\hline
Total           & 9.40 & 9.53 & 9.82 & \textbf{9.90} & \textbf{10.49} & \textbf{10.66} \\
\hline
\end{tabular}
\label{tab:ndcg}
\end{table}

\begin{table}[h]
\centering
\caption{MRR, ERR, and MAP metrics comparison}
\begin{tabular}{l|ccc|ccc}
\hline
\multirow{2}{*}{Project} & \multicolumn{3}{c|}{Rule-Based Ranking} & \multicolumn{3}{c}{LLM Ranking} \\
& MRR & ERR & MAP & MRR & ERR & MAP \\
\hline
ansible & 1.00 & 1.00 & 0.78 & 1.00 & 1.00 & \textbf{0.89} \\
cookiecutter & 0.10 & 0.41 & 0.57 & \textbf{0.20} & \textbf{0.64} & \textbf{0.76} \\
dataclasses\_json & 1.00 & 1.00 & 0.71 & 1.00 & 1.00 & \textbf{0.93} \\
docstring\_parser & 0.50 & 0.51 & \textbf{0.70} & 0.50 & 0.51 & 0.66 \\
httpie & 0.14 & 0.58 & 0.63 & \textbf{1.00} & \textbf{1.00} & \textbf{0.86} \\
isort & 0.20 & 0.24 & 0.49 & \textbf{0.33} & \textbf{0.36} & \textbf{0.56} \\
mimesis & 0.25 & 0.38 & 0.58 & 0.25 & 0.38 & 0.58 \\
pysnooper & 0.50 & 0.51 & 0.58 & 0.50 & \textbf{0.75} & \textbf{0.72} \\
sanic & 0.17 & \textbf{0.63} & \textbf{0.76} & \textbf{0.50} & 0.62 & 0.72 \\
thonny & \textbf{0.25} & 0.39 & \textbf{0.60} & 0.10 & \textbf{0.57} & 0.56 \\
tqdm & 0.50 & 0.50 & 0.56 & 0.50 & 0.50 & 0.56 \\
tornado & 0.25 & 0.44 & 0.53 & \textbf{0.33} & \textbf{0.50} & \textbf{0.69} \\
youtube\_dl & 0.25 & 0.63 & 0.56 & \textbf{1.00} & \textbf{1.00} & \textbf{0.82} \\
\hline
Total/Average & 5.11 & 7.22 & 8.05 & \textbf{7.21} & \textbf{8.83} & \textbf{9.33} \\
\hline
\end{tabular}
\label{tab:other_metrics}
\end{table}

\begin{tcolorbox}[left=0pt,right=0pt,top=0pt,bottom=0pt]
\textbf{Finding 2:} LLM-based Ranking consistently outperforms rule-based ranking across all evaluated metrics, with the advantage being most significant at top positions, which are most critical for developers who can only triage a small number of issues.

\textbf{Implication 2:} Rule-based heuristics alone are insufficient for reliably identifying the highest-priority bugs. Incorporating LLM-based re-ranking help improve the ranking, particularly for identifying the most impactful issue per project, a valued concern in resource-constrained maintenance workflow. 
\end{tcolorbox}

\subsection{Coverage-based Localization Ablation}
\label{sec:rq2-coverage}

We also investigate how test coverage level influences the quality and quantity of generated issues, to understand the contribution of the line coverage-based localization step to the overall pipeline. Projects with low test coverage naturally expose larger uncovered surfaces.
iSort (16.0\% coverage) and Tornado (30.7\% coverage) produced 24 and 284 uncovered segments, yielding 72 and 852 generated issues respectively.
Conversely, highly covered projects such as Mimesis (99.1\%) and Dataclasses-JSON (94.8\%) still produced 300 and 55 uncovered segments and 900 and 165 issues respectively, demonstrating that even near-complete test suites leave exploitable gaps that LLM-based analysis can detect.

Codebase size interacts substantially with coverage to determine issue volume.
Ansible, with 1,573 files and only 46.2\% coverage, dominates the dataset with 5,496 generated issues from 1,832 uncovered segments, over half of all generated issues.
This highlights that large codebases with moderate coverage present an important target for automated issue generation. Most projects cluster around an average of 15 to 25 lines per uncovered segment as in Table~\ref{tab:uncovered_segments}, suggesting that the uncovered regions tend to be compact, self-contained blocks rather than large untested modules.

\begin{tcolorbox}[left=0pt,right=0pt,top=0pt,bottom=0pt]
\textbf{Finding 3:} Issue volume is driven by the combined effect of
test coverage and codebase size: low-coverage, large projects (e.g.,
Ansible) generate the most issues, while even near-complete test suites
(e.g., Mimesis at 99.1\,\%) leave exploitable uncovered gaps that
contain valid bugs. Uncovered segments tend to be compact (spanning from 15 to 25 lines
on average), providing focused targets for LLM-based defect analysis.

\textbf{Implication 3:} Automated issue generation and subsequent ranking may be more impactful
for large, under tested codebases, but remains valuable even for
mature, well tested projects. Coverage percentage alone is not a
reliable indicator of residual defect risk; this suggest that developers may benefit from complementing
segment-level coverage tools (e.g., SlipCover) alongside LLM-based
analysis to identify latent bugs.
\end{tcolorbox}

\section{RQ3: Comparison with Unit Test Generation}
\label{sec:rq3}

We compare \issuespectertool{} against CoverUp~\cite{coverup}, the state-of-the-art
coverage-driven test generation tool. We run
CoverUp using the same model as \issuespectertool{} (i.e., GPT-5-mini) for a fair
comparison. To ensure a controlled comparison, we match the number of evaluated
artifacts per project: for each project, CoverUp generates $k$ unit tests over a set
of uncovered segments, while we apply the \issuespectertool{} pipeline to the same
segments and select the top-$k$ ranked issues, so that both approaches produce exactly
$k$ outputs per project. This process resulted in 168 artifacts per tool across all
evaluated projects (168 unit tests generated in total and 168 issue reports generated
in total). While RQ1 evaluates the top-10 ranked issues per project (130 issues total), this 
section adopts a matched-artifact design to ensure a fair comparison: both 
\issuespectertool{} and CoverUp are evaluated on exactly the same number of outputs 
per project, resulting in 168 artifacts per tool across all 13 projects.

\paragraph{Annotation Procedure.}
To assess the practical value of CoverUp-generated tests, we manually inspected all
tests generated for the same 168 uncovered segments analyzed by \issuespectertool{},
classifying each as Valid Bug, Requires Further Investigation, or Invalid, following
the same annotation procedure described in Section~\ref{sec:rq1}.

\paragraph{Results.}
Table~\ref{tab:coverup_comparison} summarises the results. Regarding the
CoverUp-generated tests, 33.3\% were classified as valid, 42.9\% require further
investigation, and 23.8\% were deemed invalid. Compared to \issuespectertool{}'s
actionability rate (the sum of Valid and Requiring Further Investigation categories) of 81.0\%, CoverUp's rate of 76.2\%
is similar in magnitude, though \issuespectertool{} additionally provides structured
issue reports with reproduction steps and candidate fixes, which CoverUp does not.

\begin{table}[htpb]
  \centering
  \caption{Comparison of bug assessment outcomes between CoverUp and 
  \issuespectertool{} over 168 matched artifacts per tool. Valid bugs 
  correspond to verified bugs. Issues marked as Requiring Further 
  Investigation appear plausible but require execution for confirmation 
  or represent edge cases. Invalid issues are hallucinations or 
  misunderstandings of the project requirement.}
  \label{tab:coverup_comparison}
  \begin{tabular}{lrrrr}
    \toprule
    & \multicolumn{2}{c}{\textbf{CoverUp}} 
    & \multicolumn{2}{c}{\textbf{\issuespectertool{}}} \\
    \cmidrule(lr){2-3} \cmidrule(lr){4-5}
    \textbf{Bug Assessment} & \textbf{Count} & \textbf{\%} & \textbf{Count} & \textbf{\%} \\
    \midrule
    Valid Bug                      & 56  & 33.3\% & 70  & 41.7\% \\
    Req.\ Further Inv.\            & 72  & 42.9\% & 66  & 39.3\% \\
    Invalid                        & 40  & 23.8\% & 32  & 19.0\% \\
    \midrule
    \textbf{Total}                 & \textbf{168} & \textbf{100\%} & \textbf{168} & \textbf{100\%} \\
    \bottomrule
  \end{tabular}
\end{table}


During our manual investigation of the unit tests generated by CoverUp, we note that
over-mocking occurs when tests for segments interacting with complex framework objects
(e.g.\ matplotlib canvas, Tornado async primitives) replacing these objects with stubs
that subtly diverge in behavior. In addition, invalid unit tests arise from trivial
targets and fake modules injections. Trivial targets happen when CoverUp sometimes
generates tests for simple code segments, such as a pass statement or basic attribute
assignments, where no meaningful test can exist. Our second observed issue involves
injecting fake modules into \texttt{sys.modules} and testing import mechanisms rather
than runtime logic. The 23.8\% invalid rate suggests that LLM-based test generators
struggle with trivial cases and complex object instantiation.

\begin{figure}[htpb]
  \centering
  \includegraphics[width=\linewidth]{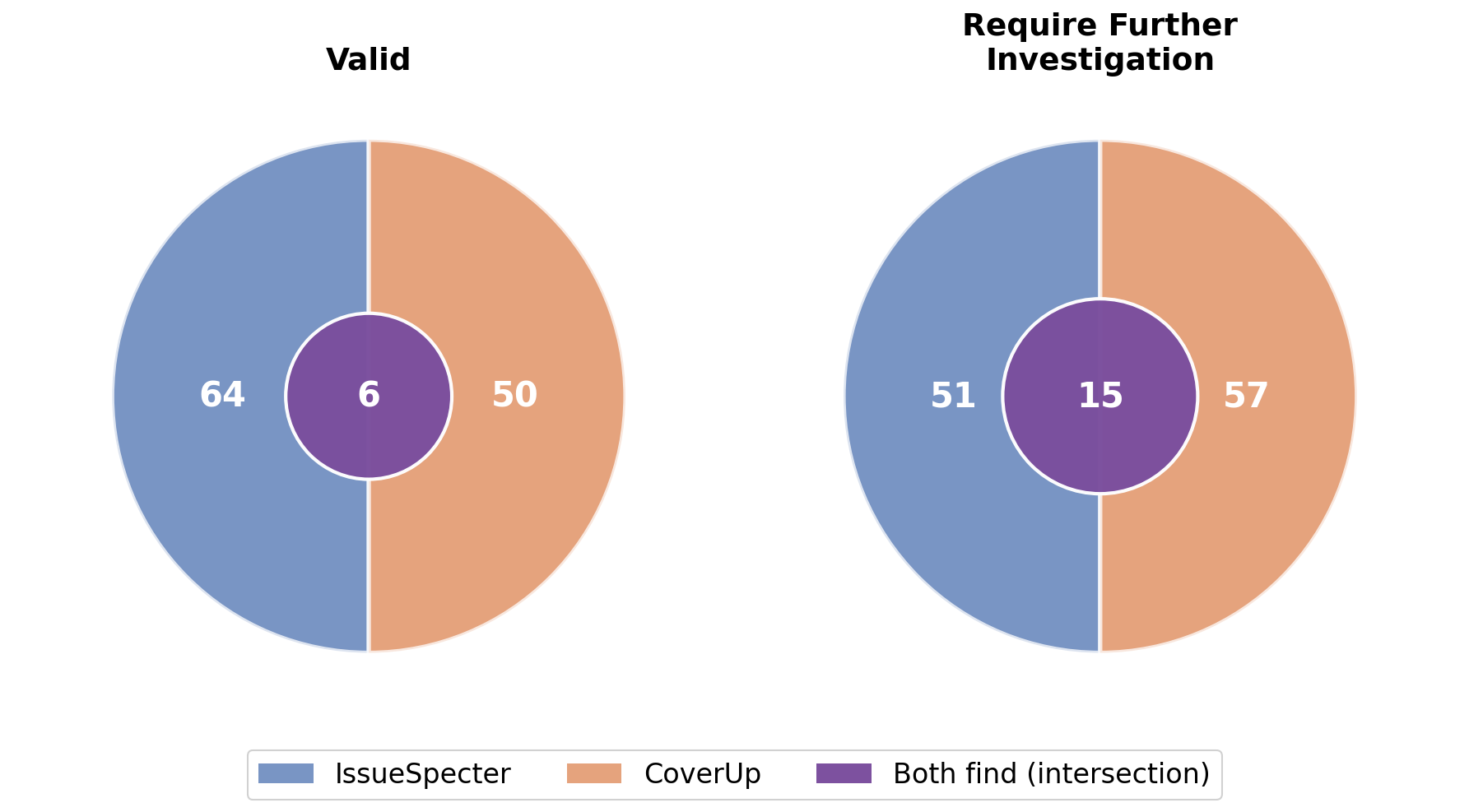}
  \caption{Overlap between bugs found by \issuespectertool{} and CoverUp 
  across 168 matched artifacts. Each circle is split into tool-exclusive 
  bugs found (for both \issuespectertool{} and CoverUp) and shared 
  bugs found (center). The left diagram represents the Valid bugs, and the right diagram accounts for  Issues requiring 
  further investigation.}
  \label{fig:overlap}
\end{figure}

Figure~\ref{fig:overlap} shows the overlap between actionable findings 
from both tools. For valid bugs, \issuespectertool{} exclusively identifies 
64 bugs while CoverUp exclusively identifies 50, with only 6 bugs found by 
both tools. For issues requiring further investigation, \issuespectertool{} 
exclusively surfaces 51 cases and CoverUp 57, with 15 shared. The low 
intersection in both categories (6 and 15 respectively) suggests that the 
two approaches are largely \emph{complementary}: each tool exposes a 
distinct set of defects, and combining both could surface a broader range 
of latent bugs than either approach alone.

\begin{tcolorbox}[left=0pt,right=0pt,top=0pt,bottom=0pt]
\textbf{Finding 4:} Both \issuespectertool{} and CoverUp achieve comparable
bug detection rates (81.0\% vs.\ 76.2\%), with \issuespectertool{} holding
a slight advantage.

\textbf{Implication 4:} Both approaches are viable for surfacing bugs
in uncovered code, but \issuespectertool{} offers a slightly lower invalid
rate and produces structured issue reports with reproduction steps and
candidate fixes that are immediately actionable, whereas interpreting the
intent and validity of a generated unit test requires additional effort from
the developer and a deeper understanding of the project's codebase.
\end{tcolorbox}

\section{Case Study} \label{sec:case_study}
We illustrate the importance and the diversity of bug detected by \issuespectertool{} by discussing three case studies.
\subsection{Case 1: Out of Memory Vulnerability in HTTP Client Library}

\definecolor{diffgreen}{rgb}{0.0, 0.5, 0.0}
\definecolor{diffred}{rgb}{0.8, 0.0, 0.0}

\begin{figure}[!t]
\begin{mdframed}[
    linewidth=0.5pt,
    innerleftmargin=6pt,
    innerrightmargin=6pt,
    innertopmargin=4pt,
    innerbottommargin=4pt
]
\begin{Verbatim}[fontsize=\scriptsize, commandchars=\\\{\}]
 class BufferedPrettyStream(PrettyStream):
     CHUNK_SIZE = 1024 * 10
\textcolor{diffgreen}{+    MAX_BUFFER_SIZE = 10 * 1024 * 1024  # 10 MiB}
     def iter_body(self) -> Iterable[bytes]:
         converter = None
         body = bytearray()
         for chunk in self.msg.iter_body(self.CHUNK_SIZE):
             if not converter and b'\textbackslash 0' in chunk:
                 converter = self.conversion.get_converter(self.mime)
                 if not converter:
                     raise BinarySuppressedError()
\textcolor{diffred}{-            body.extend(chunk)  # no size limit here}
\textcolor{diffgreen}{+            body.extend(chunk)}
\textcolor{diffgreen}{+            if len(body) > self.MAX_BUFFER_SIZE:}
\textcolor{diffgreen}{+                raise ValueError(}
\textcolor{diffgreen}{+                    f"Response body exceeds maximum buffered size of "}
\textcolor{diffgreen}{+                    f"{self.MAX_BUFFER_SIZE} bytes"}
\textcolor{diffgreen}{+                )}
         if converter:
             self.mime, body = converter.convert(body)
         yield self.process_body(body)
\end{Verbatim}
\end{mdframed}
\caption{Diff of the original and fixed \texttt{BufferedPrettyStream.iter\_body()}.}
\label{fig:oom_diff}
\end{figure}

\issuespectertool{} identified an out-of-memory vulnerability in the same HTTP 
client library from our Section \ref{section:motivating_example}, where 
\texttt{Buffered\-PrettyStream.iter\_body()} accumulates all response chunks into 
a byte array without any size validation, causing out-of-memory conditions when 
processing large responses.

\noindent\emph{Manifestation.} We reproduced the defect by crafting a synthetic 
20\,MiB response (2,000 chunks of 10\,KB each) exceeding the proposed 10\,MiB limit, verifying that a \texttt{ValueError} is raised when the limit is exceeded. The generated issue provided clear high-level reproduction steps that guided our investigation, describing how to serve a large HTTP response and observe memory exhaustion, which guided our investigation.

\noindent\emph{Root Cause.} The defect stems from an unbounded accumulation 
loop in which \texttt{body.extend(chunk)} is called unconditionally regardless 
of available system memory, with no size check or early termination.

\noindent\emph{Fix.} Figure~\ref{fig:oom_diff} presents the diff between the original code of the project
and the patch automatically generated by \issuespectertool{}, without any manual adaptation. The patch adds a configurable maximum buffer size constant and raises a descriptive \texttt{ValueError} when exceeded, allowing a graceful failure.

\noindent\emph{Classification and Impact.} Following prior work~\cite{bugpilot}, 
we classify this as a ``state consistency/ bookkeeping/caching'' bug of medium 
severity. While CLI usage causes local crashes only, the vulnerability becomes 
more severe when the library is integrated into web services or CI/CD pipelines.



\subsection{Case 2: Data Loss in Tornado's Gzip Decompressor}

\begin{figure}[!t]
\begin{mdframed}[
    linewidth=0.5pt,
    innerleftmargin=6pt,
    innerrightmargin=6pt,
    innertopmargin=4pt,
    innerbottommargin=4pt
]
\begin{Verbatim}[fontsize=\scriptsize, commandchars=\\\{\}]
 def decompress(self, value: bytes,
                max_length: int = 0) -> bytes:
     data = value
     out = bytearray()
\textcolor{diffgreen}{+    remaining = max_length}

\textcolor{diffred}{-    for chunk in iterate_chunks(data):}
\textcolor{diffred}{-        result = self.decompressobj.decompress(chunk)}
\textcolor{diffred}{-        out.extend(result)}
\textcolor{diffgreen}{+    while True:}
\textcolor{diffgreen}{+        if remaining:}
\textcolor{diffgreen}{+            chunk = self.decompressobj.decompress(}
\textcolor{diffgreen}{+                data, remaining)}
\textcolor{diffgreen}{+        else:}
\textcolor{diffgreen}{+            chunk = self.decompressobj.decompress(data)}
\textcolor{diffgreen}{+        out.extend(chunk)}
\textcolor{diffgreen}{+        if remaining:}
\textcolor{diffgreen}{+            remaining = max(0, max_length - len(out))}
\textcolor{diffgreen}{+            if remaining == 0:}
\textcolor{diffgreen}{+                break}
\textcolor{diffgreen}{+        unused = getattr(self.decompressobj,}
\textcolor{diffgreen}{+                         "unused_data", b"")}
\textcolor{diffgreen}{+        if unused:}
\textcolor{diffgreen}{+            data = unused}
\textcolor{diffgreen}{+            self.decompressobj = zlib.decompressobj(}
\textcolor{diffgreen}{+                16 + zlib.MAX_WBITS)}
\textcolor{diffgreen}{+            continue}
\textcolor{diffgreen}{+        break}

     return bytes(out)
\end{Verbatim}
\end{mdframed}
\caption{Diff of the original and fixed \texttt{GzipDecompressor.decompress()}.}
\label{fig:tornado_diff}
\end{figure}

\issuespectertool{} identified a data loss vulnerability in Tornado's gzip where \texttt{GzipDecompressor.decompress()} 
silently discards all members beyond the first in a concatenated gzip stream, 
and fails to check the \texttt{unused\_data} attribute of the zlib decompressor 
for remaining bytes.

\noindent\emph{Manifestation.} We reproduced the defect by concatenating two 
gzip-compressed segments and passing them to the decompressor. The output contained only the first segment without giving any warning. The generated issue described concrete reproduction steps: compressing two byte strings separately, concatenating the resulting gzip members, and feeding the combined stream to the decompressor. Based on the high-level logic provided in the generated report, we
manually refined these steps into an executable test. 

\noindent\emph{Root Cause.} The method assumes a single-member gzip stream and 
returns immediately after the first decompression completes---an incomplete 
state tracking assumption that fails silently when processing multi-member 
streams.

\noindent\emph{Fix.} The automatically generated patch by \issuespectertool{} shown in Figure \ref{fig:tornado_diff} introduces a loop that checks \texttt{unused\_data} 
after each decompression step, instantiating a new decompressor for each 
subsequent member until the stream is fully processed.  

\noindent\emph{Classification and Impact.} Following prior work~\cite{bugpilot}, 
we classify this as a ``state consistency/ bookkeeping/caching'' bug of medium 
severity, as it causes silent data loss in systems relying on multi-member gzip 
streams, a pattern common in HTTP streaming and log rotation tools. After submitting the 
issue to the Tornado repository, another GitHub user confirmed the 
problem\footnote{\url{https://github.com/tornadoweb/tornado/issues/3560}} and 
opened a pull request adopting our proposed 
fix\footnote{\url{https://github.com/tornadoweb/tornado/pull/3577}}.



\subsection{Case 3: Type Constraint Violation in Cookiecutter's Prompt Handler}

\begin{figure}[!t]
\begin{mdframed}[
    linewidth=0.5pt,
    innerleftmargin=6pt,
    innerrightmargin=6pt,
    innertopmargin=4pt,
    innerbottommargin=4pt
]
\begin{Verbatim}[fontsize=\scriptsize, commandchars=|\{\}]
 if prompts and var_name in prompts:
     if isinstance(prompts[var_name], str):
         question = prompts[var_name]
     else:
         if "__prompt__" in prompts[var_name]:
             question = prompts[var_name]["__prompt__"]
|textcolor{diffgreen}{+        prom_map = prompts[var_name]}
|textcolor{diffgreen}{+        def _label_for(value):}
|textcolor{diffgreen}{+            try:}
|textcolor{diffgreen}{+                label = prom_map.get(value)}
|textcolor{diffgreen}{+            except TypeError:  # value is unhashable}
|textcolor{diffgreen}{+                label = None}
|textcolor{diffgreen}{+            if label is None:}
|textcolor{diffgreen}{+                label = prom_map.get(str(value))}
|textcolor{diffgreen}{+            return label if label is not None \}
|textcolor{diffgreen}{+                         else str(value)}
|textcolor{diffred}{-        choice_lines = (}
|textcolor{diffred}{-            f"[bold magenta]{i}[/] - [bold]"}
|textcolor{diffred}{-            f"{prompts[var_name][p]}[/]"}
|textcolor{diffred}{-            if p in prompts[var_name]}
|textcolor{diffred}{-            else f"[bold magenta]{i}[/] - [bold]{p}[/]"}
|textcolor{diffred}{-            for i, p in choice_map.items())}
|textcolor{diffgreen}{+        choice_lines = (}
|textcolor{diffgreen}{+            f"[bold magenta]{i}[/] - "}
|textcolor{diffgreen}{+            f"[bold]{_label_for(value)}[/]"}
|textcolor{diffgreen}{+            for i, value in choice_map.items())}
\end{Verbatim}
\end{mdframed}
\caption{Diff of the original and fixed \texttt{read\_user\_choice()} in Cookiecutter.}
\label{fig:cookiecutter_diff}
\end{figure}

\issuespectertool{} detected a type constraint violation in Cookiecutter's 
prompt handler, where \texttt{read\_user\_choice()} raises a \texttt{TypeError} 
when option values are unhashable types (e.g., dictionaries or lists), as they 
are used directly as dictionary lookup keys.

\noindent\emph{Manifestation.} We reproduced the defect by constructing a template 
with nested dictionary cloud configuration options, triggering a 
\texttt{TypeError: unhashable type: 'dict'} crash that halted template 
generation, with the guidance of the high-level reproduction steps generated by our tool. The generated issue included concrete reproduction steps showing exactly how to construct a prompts mapping with string-keyed representations of dict options, making it straightforward to trigger and confirm the crash with minimal manual effort. 

\noindent\emph{Root Cause.} The handler implicitly assumes all option values are 
hashable, violating Cookiecutter's own support for arbitrary JSON structures 
as valid option values (a latent inconsistency undetected by static type checkers).

\noindent\emph{Fix.} The automatically generated patch by \issuespectertool{} in Figure \ref{fig:cookiecutter_diff} introduces a \texttt{\_label\_for()} helper that catches 
\texttt{TypeError} for unhashable values and falls back to their string 
representation as the lookup key.
\noindent\emph{Classification and Impact.} Following prior work~\cite{bugpilot}, we 
classify this as a type handling bug of medium severity, as it prevents users 
from defining templates with structured configuration options.



\section{Related Work}

\noindent\textbf{Bug Report Prioritization.} 

Bug report prioritization has been
extensively studied as a classification problem, with early work establishing the
core feature sets and classifiers that remain relevant today. Our rule-based criteria is derived from prior study
on priority prediction
~\cite{preranking_criteria}
which investigates the effect of two feature
sets (i.e., textual content drawn from bug report summaries and structured metadata
such as component, operating system, and severity) on classification accuracy
using Naive Bayes, Decision Trees, and Random Forest across Eclipse and Firefox
Bugzilla datasets. Results show that metadata-based features substantially
outperform text-only features, and that Random Forest and Decision Tree
consistently surpass Naive Bayes. Building on this, a multi-factor approach known
as DRONE~\cite{tian2013drone} proposes an automated priority recommendation
framework that extracts temporal, textual, author, related-report, severity, and
product features from bug reports, training a discriminative model that handles
ordinal class labels and imbalanced data. Evaluated on over 100,000 Eclipse bug
reports, DRONE outperforms severity-based baselines in average F-measure,
particularly for high-priority reports. Subsequent work has explored richer signal
sources for prioritization: an emotion-based approach~\cite{umer2018emotion} augments
DRONE-style feature vectors with affect scores derived from the natural language of
bug descriptions, while a deep learning study~\cite{bani2021deep} employs RNN-LSTM to
improve F-measure over SVM and KNN baselines on Eclipse and Mozilla. More recently,
transformer-based approaches have pushed the state of the art further. Meanwhile, a
Convolutional Neural Network (CNN)-based study~\cite{umer2019cnn} combines topic modeling via a variant of
LDA with text classification on 85,156 Eclipse Platform bug reports, outperforming
DRONE and other prior approaches in accuracy, precision, and recall. 
In contrast to these
data-driven approaches, which require large labeled corpora of historically
prioritized reports, \issuespectertool{} assigns severity ratings to newly
generated issue reports using a rule-based technique applied at defect discovery
time, without relying on any historical priority labels or training data.

\noindent\textbf{GitHub Issue Resolution.} Prior work ~\cite{ehsani} empirically investigates what makes developer-ChatGPT conversations effective for software issue resolution, analysing 686 conversations shared within GitHub issue threads. They find that only 62\% of these conversations are helpful, with ChatGPT performing best on well-scoped tasks such as code generation and tool/library recommendations, but struggling with complex, project-specific problems such as system-level debugging and refactoring. MAGIS~\cite{magis} proposes a multi-agent LLM framework for resolving GitHub 
issues, comprising four specialized agents: a Manager, Repository Custodian, 
Developer, and Quality Assurance Engineer. Evaluated on SWE-bench, MAGIS 
resolves 13.94\% of issues, an eight-fold improvement over direct GPT-4 
application. Unlike \issuespectertool{}, which focuses on surfacing latent 
bugs in uncovered code and producing actionable reports, MAGIS targets 
repository-level patch generation for existing reported issues. SWE-Fixer~\cite{xie2025swefixer} introduces an open-source pipeline-based framework 
for GitHub issue resolution, decomposing the task into coarse-to-fine file 
retrieval (BM25 combined with a fine-tuned 7B model) and code editing (a 72B 
model trained on 110K curated instances). Evaluated on SWE-Bench, SWE-Fixer 
achieves 22.0\% on Lite and 30.2\% on Verified, requiring only two model calls 
per instance. Unlike \issuespectertool{}, which proactively surfaces latent bugs 
in uncovered code, SWE-Fixer targets patch generation for already-reported issues.

\noindent\textbf{LLM-based Technique on Bug Report Enhancement.} Prior work on automated bug report enhancement~\cite{acharya2025} proposes an approach to automatically transform unstructured bug reports into high-quality, structured ones, evaluating instruction fine-tuned LLMs (Qwen~2.5, Mistral, and Llama~3.2) against Chat-GPT-4o using the CTQRS (Crowdsourced Test Report Quality Score) criteria~\cite{Zhang_CQTRS}, a framework that systematically scores bug reports by combining morphological, relational, and analytical indicators through dependency parsing. Fine-tuned on 3,966 high-quality Bugzilla reports, the best-performing model achieves 77\% CTQRS, outperforming other open-source models and matching ChatGPT-4o, while also demonstrating cross-project generalizability on unseen ecosystems such as Eclipse and GCC. Building on this foundation, a subsequent multi-agent approach proposes AgentReport~\cite{choi_agentreport}, an LLM pipeline that addresses the modularity, reproducibility, and scalability limitations of single-pipeline approaches. AgentReport integrates QLoRA-4bit fine-tuning, CTQRS-based structured prompting, Chain-of-Thought reasoning, and FAISS-based one-shot exemplar retrieval. Evaluated on the same 3,966 Bugzilla pairs, it achieves 80.5\% CTQRS, 84.6\% ROUGE-1 Recall, and 86.4\% SBERT, outperforming GPT-4o in both zero-shot and few-shot settings. While these studies focus on improving the quality of bug reports that developers write, \issuespectertool{} operates upstream (i.e., it automatically detects latent defects in uncovered code and generates issue reports without requiring any existing bug report corpus).

\noindent\textbf{LLM-based Test Generation.} Several test generation techniques have been proposed to generate unit test automatically~\cite{coverup,codamosa,schafer, unit_test_cov}.
CodaMosa~\cite{codamosa} was proposed to address coverage plateaus in traditional SBST approaches. CodaMosa monitors Pynguin's~\cite{pynguin} search progress and, when coverage stalls, queries Codex to generate test cases targeting under-covered functions, which are then reintegrated into the search. Previous work \cite{unit_test_cov} proposes fine-tuning \texttt{DeepSeek-R1-Distill-Qwen-7B} with AST-structured prompts and iterative coverage feedback for automated unit test generation. Evaluated on seven open-source projects, it achieves up to 88.6\% line coverage, 83.3\% branch coverage, and a 78.9\% mutation score, outperforming both the base model by over 20\% and Pynguin.

CoverUp~\cite{coverup} is a LLM-based test generation for Python that measures existing test suite coverage using SlipCover~\cite{slipcover}, identifies functions and methods lacking line or branch coverage via AST analysis, and iteratively prompts an LLM with focused excerpts of the uncovered code, embedding precise coverage information and allowing the model to request additional context. If generated tests fail or do not improve coverage, the approach continues the dialogue with targeted feedback before accepting any test. 
  \issuespectertool{} is grounded in the same insight that LLMs can be effectively directed toward uncovered code when given precise coverage signals,  but it re-purposes coverage analysis not to generate tests, but to identify uncovered segments most likely to harbor undiscovered bugs, tasking an LLM with detecting and reporting those defects rather than writing tests to trigger them. Instead of test generation that aims to improve coverage only, \issuespectertool{} focuses on detecting latent bugs and generate structured issue reports with severity ratings, reproduction steps, and suggested fixes.
\section{Threats to validity}

To minimize bias during ranking and bug type annotation, two annotators resolved disagreements through discussion. The LLM used for issue generation and ranking (GPT-5-mini) may produce inconsistent outputs across runs due to its non-deterministic nature. The rule-based ranking relies on heuristics such as word count as a proxy for completeness, which may not always reflect true issue quality. Bugs flagged per segment are also limited to three by our prompt design, potentially causing some defects to go unreported. Generated fixes are verified against the existing test suite, but passing tests does not guarantee correctness, as a fix may still be semantically incorrect.
Furthermore, the LLM may exhibit confirmation bias toward identifying defects, tending to flag code segments as buggy even when no real defect exists, which can inflate false positives among top-ranked issues.

Finally, our evaluation is restricted to Python projects, which limits the generalizability of our findings to other programming languages. Language-specific constructs, ecosystems, and testing practices may influence both the types of bugs detected and the effectiveness of LLM-based analysis. By fixing the programming language while varying project domains, we isolate the effect of domain diversity on bug detection without introducing language-related features that would complicate interpretation of results.

\section{Conclusions}
We present \issuespectertool{}, an  automated pipeline that transforms uncovered code segments into ranked, actionable issue reports by combining coverage-guided localization with LLM-based defect analysis and two-stage ranking. Evaluated on 13 open-source Python projects, \textsc{IssueSpecter} generated \totalissues issue reports, of which 84.6\% of the top-ranked issues were either confirmed valid bugs (37.7\%) or warranted further investigation (46.9\%), with only 15.4\% classified as false positives. LLM-based ranking consistently outperformed rule-based ranking across all metrics, improving MRR by 41\% and P@3 by 50\%. Our case studies confirmed the practical value of our tool, surfacing a path traversal vulnerability (CWE-22), a silent data-loss bug in a gzip decompressor, and a type constraint violation. Compared to CoverUp, our adopted baseline, \issuespectertool{} achieves a slightly superior comparable actionability rate  while additionally providing structured reports with reproduction steps and proposed Python fixes. These results suggest that LLM-based analysis of uncovered code is a practical complement to existing test suites for surfacing probable latent diverse bugs across Python projects from distinct domains.

\bibliographystyle{plain}

\bibliography{sigproc}

@software{pysnooper,
    title={PySnooper: Never use print for debugging again},
    author={Rachum, Ram and Hall, Alex and Yanokura, Iori and others},
    year={2019},
    month={jun},
    publisher={PyCon Israel},
    doi={10.5281/zenodo.10462459},
    url={https://github.com/cool-RR/PySnooper}
}

@software{tqdm,
       author = {{da Costa-Luis}, Casper and {Larroque}, Stephen Karl and {Altendorf}, Kyle and {Mary}, Hadrien and {richardsheridan} and {Korobov}, Mikhail and {Yorav-Raphael}, Noam and {Ivanov}, Ivan and {Bargull}, Marcel and {Rodrigues}, Nishant and {Shawn} and {Dektyarev}, Mikhail and {G{\'o}rny}, Micha{\l} and {mjstevens777} and {Pagel}, Matthew D. and {Zugnoni}, Martin and {JC} and {CrazyPython} and {Newey}, Charles and {Lee}, Antony and {pgajdos} and {Todd} and {Malmgren}, Staffan and {redbug312} and {Desh}, Orivej and {Nechaev}, Nikolay and {Boyle}, Mike and {Nordlund}, Max and {MapleCCC} and {McCracken}, Jack},
        title = "{tqdm: A fast, Extensible Progress Bar for Python and CLI}",
         year = 2024,
        month = nov,
          eid = {10.5281/zenodo.595120},
          doi = {10.5281/zenodo.595120},
      version = {v4.67.1},
    publisher = {Zenodo},
       adsurl = {https://ui.adsabs.harvard.edu/abs/2022zndo....595120D}
}

@inproceedings{thonny,
author = {Annamaa, Aivar},
title = {Introducing Thonny, a Python IDE for learning programming},
year = {2015},
isbn = {9781450340205},
publisher = {Association for Computing Machinery},
address = {New York, NY, USA},
url = {https://doi.org/10.1145/2828959.2828969},
doi = {10.1145/2828959.2828969},
booktitle = {Proceedings of the 15th Koli Calling Conference on Computing Education Research},
pages = {117–121},
numpages = {5},
location = {Koli, Finland},
series = {Koli Calling '15}
}

@software{ansible,
    title = {Ansible: Simple IT Automation},
    author = {{Michael DeHaan} and {Ansible Project Contributors}},
    url = {https://github.com/ansible/ansible}
}

@software{cookiecutter,
    title = {Cookiecutter: A Command-Line Utility That Creates Projects from Project Templates},
    author = {{Audrey M. Roy Greenfeld} and {Cookiecutter Contributors}},
    url = {https://github.com/cookiecutter/cookiecutter}
}

@software{dataclasses_json,
    title = {dataclasses-json: A Library for Serializing Python Dataclasses to JSON},
    author = {{dataclasses-json Contributors}},
    url = {https://github.com/lidatong/dataclasses-json}
}

@software{docstring_parser,
    title = {docstring-parser: A Python Library to Parse Docstrings},
    author = {{docstring-parser Contributors}},
    url = {https://github.com/rr-/docstring_parser}
}

@software{httpie,
    title = {HTTPie: a command-line HTTP client},
    author = {{HTTPie Contributors}},
    url = {https://github.com/httpie/cli}
}

@software{isort,
    title = {isort: A Python Utility / Library to Sort Imports},
    author = {{Timothy Crosley} and {isort Contributors}},
    url = {https://github.com/PyCQA/isort}
}

@software{mimesis,
    title = {Mimesis: A Python Library for Generating Fake Data},
    author = {{Mimesis Contributors}},
    url = {https://github.com/lk-geimfari/mimesis}
}

@software{sanic,
    title = {Sanic Framework},
    author = {{Adam Hopkins} and {Sanic Contributors}},
    url = {https://github.com/sanic-org/sanic}
}

@software{tornado,
    title = {Tornado: A Python Web Framework and Asynchronous Networking Library},
    author = {{Tornado Contributors}},
    url = {https://github.com/tornadoweb/tornado}
}

@software{youtube_dl,
    title = {youtube-dl: A Command-Line Program to Download Videos from YouTube and Other Sites},
    author = {{Daniel Bolton} and {youtube-dl Contributors}},
    url = {https://github.com/ytdl-org/youtube-dl}
}

@INPROCEEDINGS{multi_agents_issue,
  author={Adapa, Chathurya and A R K, Anjana and Rahim, Rafsal and Victor, Ajay},
  booktitle={2025 IEEE International Conference for Women in Innovation, Technology \& Entrepreneurship (ICWITE)},
  title={A Multi-Agent AI Framework for Agile Workflow Automation, Issue Resolution, and Developer Performance Evaluation}, 
  year={2025},
  volume={},
  number={},
  pages={1-6},
  doi={10.1109/ICWITE64848.2025.11306978}}

@inproceedings{tian2013drone,
  title={Drone: Predicting priority of reported bugs by multi-factor analysis},
  author={Tian, Yuan and Lo, David and Sun, Chengnian},
  booktitle={2013 IEEE International Conference on Software Maintenance},
  pages={200--209},
  year={2013},
  organization={IEEE}
}

@article{umer2019cnn,
  title={CNN-based automatic prioritization of bug reports},
  author={Umer, Qasim and Liu, Hui and Illahi, Inam},
  journal={IEEE Transactions on Reliability},
  volume={69},
  number={4},
  pages={1341--1354},
  year={2019},
  publisher={IEEE}
}

@article{bani2021deep,
  title={A deep-learning-based bug priority prediction using RNN-LSTM neural networks},
  author={Bani-Salameh, Hani and Sallam, Mohammed and Al shboul, Bashar},
  journal={e-Informatica Software Engineering Journal},
  volume={15},
  number={1},
  pages={29--45},
  year={2021},
  publisher={Politechnika Wroc{\l}awska. Oficyna Wydawnicza Politechniki Wroc{\l}awskiej}
}

@article{umer2018emotion,
  title={Emotion based automated priority prediction for bug reports},
  author={Umer, Qasim and Liu, Hui and Sultan, Yasir},
  journal={IEEE Access},
  volume={6},
  pages={35743--35752},
  year={2018},
  publisher={IEEE}
}

@inproceedings{magis,
 author = {Tao, Wei and Zhou, Yucheng and Wang, Yanlin and Zhang, Wenqiang and Zhang, Hongyu and Cheng, Yu},
 booktitle = {Advances in Neural Information Processing Systems},
 doi = {10.52202/079017-1647},
 editor = {A. Globerson and L. Mackey and D. Belgrave and A. Fan and U. Paquet and J. Tomczak and C. Zhang},
 pages = {51963--51993},
 publisher = {Curran Associates, Inc.},
 title = {MAGIS: LLM-Based Multi-Agent Framework for GitHub Issue Resolution},
 url = {https://proceedings.neurips.cc/paper_files/paper/2024/file/5d1f02132ef51602adf07000ca5b6138-Paper-Conference.pdf},
 volume = {37},
 year = {2024}
}

@misc{chen2026sweexpexperience,
      title={SWE-Exp: Experience-Driven Software Issue Resolution}, 
      author={Silin Chen and Shaoxin Lin and Yuling Shi and Heng Lian and Xiaodong Gu and Longfei Yun and Dong Chen and Lin Cao and Jiyang Liu and Nu Xia and Qianxiang Wang},
      year={2026},
      eprint={2507.23361},
      archivePrefix={arXiv},
      primaryClass={cs.SE},
      url={https://arxiv.org/abs/2507.23361}, 
}

@misc{li2026llmbasedissue,
      title={Advances and Frontiers of LLM-based Issue Resolution in Software Engineering: A Comprehensive Survey}, 
      author={Caihua Li and Lianghong Guo and Yanlin Wang and Daya Guo and Wei Tao and Zhenyu Shan and Mingwei Liu and Jiachi Chen and Haoyu Song and Duyu Tang and Hongyu Zhang and Zibin Zheng},
      year={2026},
      eprint={2601.11655},
      archivePrefix={arXiv},
      primaryClass={cs.SE},
      url={https://arxiv.org/abs/2601.11655}, 
}

@article{ehsani,
author = {Ehsani, Ramtin and Pathak, Sakshi and Parra, Esteban and Haiduc, Sonia and Chatterjee, Preetha},
year = {2025},
month = {11},
pages = {},
title = {What characteristics make ChatGPT effective for software issue resolution? An empirical study of task, project, and conversational signals in GitHub issues},
volume = {31},
journal = {Empirical Software Engineering},
doi = {10.1007/s10664-025-10745-8}
}

@inproceedings{xie2025swefixer,
    title = "{SWE}-Fixer: Training Open-Source {LLM}s for Effective and Efficient {G}it{H}ub Issue Resolution",
    author = "Xie, Chengxing  and
      Li, Bowen  and
      Gao, Chang  and
      Du, He  and
      Lam, Wai  and
      Zou, Difan  and
      Chen, Kai",
    editor = "Che, Wanxiang  and
      Nabende, Joyce  and
      Shutova, Ekaterina  and
      Pilehvar, Mohammad Taher",
    booktitle = "Findings of the Association for Computational Linguistics: ACL 2025",
    month = jul,
    year = "2025",
    address = "Vienna, Austria",
    publisher = "Association for Computational Linguistics",
    url = "https://aclanthology.org/2025.findings-acl.62/",
    doi = "10.18653/v1/2025.findings-acl.62",
    pages = "1123--1139",
    ISBN = "979-8-89176-256-5"
}

@article{pan2024swegym,
  title={Training Software Engineering Agents and Verifiers with SWE-Gym},
  author={Jiayi Pan and Xingyao Wang and Graham Neubig and Navdeep Jaitly and Heng Ji and Alane Suhr and Yizhe Zhang},
  journal={ArXiv},
  year={2024},
  volume={abs/2412.21139},
  url={https://api.semanticscholar.org/CorpusID:275133330}
}

@Article{choi_agentreport,
AUTHOR = {Choi, Seojin and Yang, Geunseok},
TITLE = {AgentReport: A Multi-Agent LLM Approach for Automated and Reproducible Bug Report Generation},
JOURNAL = {Applied Sciences},
VOLUME = {15},
YEAR = {2025},
NUMBER = {22},
ARTICLE-NUMBER = {11931},
URL = {https://www.mdpi.com/2076-3417/15/22/11931},
ISSN = {2076-3417},
DOI = {10.3390/app152211931}
}

@ARTICLE{schafer,
  author={Schäfer, Max and Nadi, Sarah and Eghbali, Aryaz and Tip, Frank},
  journal={IEEE Transactions on Software Engineering}, 
  title={An Empirical Evaluation of Using Large Language Models for Automated Unit Test Generation}, 
  year={2024},
  volume={50},
  number={1},
  pages={85-105},
  doi={10.1109/TSE.2023.3334955}}

@inproceedings{acharya2025,
author = {Acharya, Jagrit and Ginde, Gouri},
title = {Can We Enhance Bug Report Quality Using LLMs?: An Empirical Study of LLM-Based Bug Report Generation},
year = {2025},
isbn = {9798400713859},
publisher = {Association for Computing Machinery},
address = {New York, NY, USA},
url = {https://doi.org/10.1145/3756681.3756995},
doi = {10.1145/3756681.3756995},
booktitle = {Proceedings of the 29th International Conference on Evaluation and Assessment in Software Engineering},
pages = {994–1003},
numpages = {10},
location = {
},
series = {EASE '25}
}

@INPROCEEDINGS{Zhang_CQTRS,
  author={Zhang, Huan and Zhao, Yuan and Yu, Shengcheng and Chen, Zhenyu},
  booktitle={2022 9th International Conference on Dependable Systems and Their Applications (DSA)}, 
  title={Automated Quality Assessment for Crowdsourced Test Reports Based on Dependency Parsing}, 
  year={2022},
  volume={},
  number={},
  pages={34-41},
  doi={10.1109/DSA56465.2022.00014}}

@article{chatgpt_units,
author = {Yuan, Zhiqiang and Liu, Mingwei and Ding, Shiji and Wang, Kaixin and Chen, Yixuan and Peng, Xin and Lou, Yiling},
title = {Evaluating and Improving ChatGPT for Unit Test Generation},
year = {2024},
issue_date = {July 2024},
publisher = {Association for Computing Machinery},
address = {New York, NY, USA},
volume = {1},
number = {FSE},
url = {https://doi.org/10.1145/3660783},
doi = {10.1145/3660783},
journal = {Proc. ACM Softw. Eng.},
month = jul,
articleno = {76},
numpages = {24}
}

@article{pynguin_extension,
author = {Lukasczyk, Stephan and Kroiß, Florian and Fraser, Gordon},
year = {2023},
month = {01},
pages = {},
title = {An empirical study of automated unit test generation for Python},
volume = {28},
journal = {Empirical Software Engineering},
doi = {10.1007/s10664-022-10248-w}
}

@inproceedings{gpt_pynguin_units_py,
author = {Bhatia, Shreya and Gandhi, Tarushi and Kumar, Dhruv and Jalote, Pankaj},
title = {Unit Test Generation using Generative AI : A Comparative Performance Analysis of Autogeneration Tools},
year = {2024},
isbn = {9798400705793},
publisher = {Association for Computing Machinery},
address = {New York, NY, USA},
url = {https://doi.org/10.1145/3643795.3648396},
doi = {10.1145/3643795.3648396},
booktitle = {Proceedings of the 1st International Workshop on Large Language Models for Code},
pages = {54–61},
numpages = {8},
location = {Lisbon, Portugal},
series = {LLM4Code '24}
}

@INPROCEEDINGS{gen_ai_python,
  author={Jiri, Medlen and Emese, Bari and Medlen, Patrick},
  booktitle={2024 IEEE International Conference on Artificial Intelligence Testing (AITest)}, 
  title={Leveraging Large Language Models for Python Unit Test}, 
  year={2024},
  volume={},
  number={},
  pages={95-100},
  doi={10.1109/AITest62860.2024.00020}}

@misc{pynguin_type_inf,
      title={Combining Type Inference and Automated Unit Test Generation for Python}, 
      author={Lukas Krodinger and Stephan Lukasczyk and Gordon Fraser},
      year={2025},
      eprint={2507.01477},
      archivePrefix={arXiv},
      primaryClass={cs.SE},
      url={https://arxiv.org/abs/2507.01477}, 
}

@inproceedings{copilot_unit_test_py,
author = {El Haji, Khalid and Brandt, Carolin and Zaidman, Andy},
title = {Using GitHub Copilot for Test Generation in Python: An Empirical Study},
year = {2024},
isbn = {9798400705885},
publisher = {Association for Computing Machinery},
address = {New York, NY, USA},
url = {https://doi.org/10.1145/3644032.3644443},
doi = {10.1145/3644032.3644443},
booktitle = {Proceedings of the 5th ACM/IEEE International Conference on Automation of Software Test (AST 2024)},
pages = {45–55},
numpages = {11},
location = {Lisbon, Portugal},
series = {AST '24}
}

@INPROCEEDINGS{unit_test_cov,
  author={Long, Jianlin and Qin, Renchao and Jiang, Zujun and Duan, Jialin and Li, Suonan and Qu, Xiaosheng},
  booktitle={2025 18th International Congress on Image and Signal Processing, BioMedical Engineering and Informatics (CISP-BMEI)}, 
  title={A Python Unit Test Generation Method Based on Fine-Tuned Language Models and Coverage}, 
  year={2025},
  volume={},
  number={},
  pages={1-6},
  doi={10.1109/CISP-BMEI68103.2025.11259411}}

@inproceedings{testgenllm,
author = {Alshahwan, Nadia and Chheda, Jubin and Finogenova, Anastasia and Gokkaya, Beliz and Harman, Mark and Harper, Inna and Marginean, Alexandru and Sengupta, Shubho and Wang, Eddy},
title = {Automated Unit Test Improvement using Large Language Models at Meta},
year = {2024},
isbn = {9798400706585},
publisher = {Association for Computing Machinery},
address = {New York, NY, USA},
url = {https://doi.org/10.1145/3663529.3663839},
doi = {10.1145/3663529.3663839},
booktitle = {Companion Proceedings of the 32nd ACM International Conference on the Foundations of Software Engineering},
pages = {185–196},
numpages = {12},
location = {Porto de Galinhas, Brazil},
series = {FSE 2024}
}

@article{mutap,
title = {Effective test generation using pre-trained Large Language Models and mutation testing},
journal = {Information and Software Technology},
volume = {171},
pages = {107468},
year = {2024},
issn = {0950-5849},
doi = {https://doi.org/10.1016/j.infsof.2024.107468},
url = {https://www.sciencedirect.com/science/article/pii/S0950584924000739},
author = {Arghavan Moradi Dakhel and Amin Nikanjam and Vahid Majdinasab and Foutse Khomh and Michel C. Desmarais}
}

@inproceedings{pynguin,
author = {Lukasczyk, Stephan and Fraser, Gordon},
title = {Pynguin: automated unit test generation for Python},
year = {2022},
isbn = {9781450392235},
publisher = {Association for Computing Machinery},
address = {New York, NY, USA},
url = {https://doi.org/10.1145/3510454.3516829},
doi = {10.1145/3510454.3516829},
booktitle = {Proceedings of the ACM/IEEE 44th International Conference on Software Engineering: Companion Proceedings},
pages = {168–172},
numpages = {5},
location = {Pittsburgh, Pennsylvania},
series = {ICSE '22}
}

@INPROCEEDINGS{codamosa,
  author={Lemieux, Caroline and Inala, Jeevana Priya and Lahiri, Shuvendu K. and Sen, Siddhartha},
  booktitle={2023 IEEE/ACM 45th International Conference on Software Engineering (ICSE)}, 
  title={CodaMosa: Escaping Coverage Plateaus in Test Generation with Pre-trained Large Language Models}, 
  year={2023},
  volume={},
  number={},
  pages={919-931},
  doi={10.1109/ICSE48619.2023.00085}}

@article{coverup,
author = {Altmayer Pizzorno, Juan and Berger, Emery D.},
title = {CoverUp: Effective High Coverage Test Generation for Python},
year = {2025},
issue_date = {July 2025},
publisher = {Association for Computing Machinery},
address = {New York, NY, USA},
volume = {2},
number = {FSE},
url = {https://doi.org/10.1145/3729398},
doi = {10.1145/3729398},
journal = {Proc. ACM Softw. Eng.},
month = jun,
articleno = {FSE128},
numpages = {23}
}

@inproceedings{slipcover,
author = {Altmayer Pizzorno, Juan and Berger, Emery D.},
title = {SlipCover: Near Zero-Overhead Code Coverage for Python},
year = {2023},
isbn = {9798400702211},
publisher = {Association for Computing Machinery},
address = {New York, NY, USA},
url = {https://doi.org/10.1145/3597926.3598128},
doi = {10.1145/3597926.3598128},
booktitle = {Proceedings of the 32nd ACM SIGSOFT International Symposium on Software Testing and Analysis},
pages = {1195–1206},
numpages = {12},
location = {Seattle, WA, USA},
series = {ISSTA 2023}
}

@article{bugpilot,
  title={Bugpilot: Complex bug generation for efficient learning of swe skills},
  author={Sonwane, Atharv and White, Isadora and Lee, Hyunji and Pereira, Matheus and Caccia, Lucas and Kim, Minseon and Shi, Zhengyan and Singh, Chinmay and Sordoni, Alessandro and C{\^o}t{\'e}, Marc-Alexandre and others},
  journal={arXiv preprint arXiv:2510.19898},
  year={2025}
}

@INPROCEEDINGS{preranking_criteria,
  author={Alenezi, Mamdouh and Banitaan, Shadi},
  booktitle={2013 12th International Conference on Machine Learning and Applications}, 
  title={Bug Reports Prioritization: Which Features and Classifier to Use?}, 
  year={2013},
  volume={2},
  number={},
  pages={112-116},
  doi={10.1109/ICMLA.2013.114}}

@article{dataset,
  author = {Javaria Imtiaz and Muhammad Zohaib Iqbal and et al.},
  title = {An automated model-based approach to repair test suites of evolving web applications},
  year = {2021},
  volume = {171},
  pages = {110841},
url = {https://doi.org/10.1016/j.jss.2020.110841},
  journal = {Journal of Systems and Software}
}

@misc{reproducibility,
  author={Zhuolin Xu},
  title={ChatGPT UI Test Repair Repository},
  howpublished = {\url{https://github.com/xuzhi2021/ChatGPTUITestRepair}},
  year={2025}
}

\end{document}